% PLAIN TEX!

 % MACROS

%
% TEXT
%

% fonts

\def\famname{
 \textfont0=\textrm \scriptfont0=\scriptrm
 \scriptscriptfont0=\sscriptrm
 \textfont1=\textmi \scriptfont1=\scriptmi
 \scriptscriptfont1=\sscriptmi
 \textfont2=\textsy \scriptfont2=\scriptsy \scriptscriptfont2=\sscriptsy
 \textfont3=\textex \scriptfont3=\textex \scriptscriptfont3=\textex
 \textfont4=\textbf \scriptfont4=\scriptbf \scriptscriptfont4=\sscriptbf
 \skewchar\textmi='177 \skewchar\scriptmi='177
 \skewchar\sscriptmi='177
 \skewchar\textsy='60 \skewchar\scriptsy='60
 \skewchar\sscriptsy='60
 \def\rm{\fam0 \textrm} \def\bf{\fam4 \textbf}}
\def\sca#1{scaled\magstep#1} \def\scah{scaled\magstephalf} 
\def\twelvepoint{
 \font\textrm=cmr12 \font\scriptrm=cmr8 \font\sscriptrm=cmr6
 \font\textmi=cmmi12 \font\scriptmi=cmmi8 \font\sscriptmi=cmmi6 
 \font\textsy=cmsy10 \sca1 \font\scriptsy=cmsy8
 \font\sscriptsy=cmsy6
 \font\textex=cmex10 \sca1
 \font\textbf=cmbx12 \font\scriptbf=cmbx8 \font\sscriptbf=cmbx6
 \font\it=cmti12
 \font\sectfont=cmbx12 \sca1
 \font\sectmath=cmmib10 \sca2
 \font\sectsymb=cmbsy10 \sca2
 \font\refrm=cmr10 \scah \font\refit=cmti10 \scah
 \font\refbf=cmbx10 \scah
 \def\twelverm{\textrm} \def\twelveit{\it} \def\twelvebf{\textbf}
 \famname \textrm 
 \advance\voffset by .06in \advance\hoffset by .28in
 \normalbaselineskip=17.5pt plus 1pt \baselineskip=\normalbaselineskip
 \parindent=21pt
 \setbox\strutbox=\hbox{\vrule height10.5pt depth4pt width0pt}}

% primes and other "@"garbage

\catcode`@=11

{\catcode`\'=\active \def'{{}^\bgroup\prim@s}}

\def\screwcount{\alloc@0\count\countdef\insc@unt}   % for stupid
\def\screwdimen{\alloc@1\dimen\dimendef\insc@unt} % \outer errors
\def\screwbox{\alloc@4\box\chardef\insc@unt}

\catcode`@=12

% Text style parameters ("textile"?)

\overfullrule=0pt			% gets rid of stupid black boxes
\vsize=9in \hsize=6in
%\parskip=\medskipamount	% space between paragraphs
\lineskip=0pt				% minimum box separation
\abovedisplayskip=1.2em plus.3em minus.9em % space above equation
\belowdisplayskip=1.2em plus.3em minus.9em	% " below
\abovedisplayshortskip=0em plus.3em	% " above when no overlap
\belowdisplayshortskip=.7em plus.3em minus.4em	% " below
\parindent=21pt
\setbox\strutbox=\hbox{\vrule height10.5pt depth4pt width0pt}
\def\makefootline{\baselineskip=30pt \line{\the\footline}}
\footline={\ifnum\count0=1 \hfil \else\hss\twelverm\folio\hss \fi}
\pageno=1

% Box at relative coordinates (x,y)

\def\put(#1,#2)#3{\screwdimen\unit  \unit=1in
	\vbox to0pt{\kern-#2\unit\hbox{\kern#1\unit
	\vbox{#3}}\vss}\nointerlineskip}

% Lines & stuff

\def\\{\hfil\break}
\def\newpage{\vfill\eject}
\def\center{\leftskip=0pt plus 1fill \rightskip=\leftskip \parindent=0pt
 \def\textindent##1{\par\hangindent21pt\footrm\noindent\hskip21pt
 \llap{##1\enspace}\ignorespaces}\par}
 % use as {\center ... \par}, shorten lines with \\
\def\unnarrower{\leftskip=0pt \rightskip=\leftskip}

%% Page and section headings and reference stuff (also, \refs below)

%\def\sect#1\par{\par\ifdim\lastskip<\medskipamount
%	\bigskip\medskip\goodbreak\else\nobreak\fi
%	\noindent{\sectfont{#1}}\par\nobreak\medskip} % see Ü below 
%\def\itemize#1 {\item{[#1]}}	% see £ below; also use \itemitem 
\def\vol#1 {{\refbf#1} }		 % see É below
%\def\topic{\par\noindent \hangafter1 \hangindent20pt}
%\def\topic{\par\noindent \hangafter1 \hangindent60pt}
%\def\ref#1{\setbox0=\hbox{M}$\vbox to\ht0{}^{#1}$}

% Journal abbreviations

\def\NP #1 {{\refit Nucl. Phys.} {\refbf B{#1}} }
\def\PL #1 {{\refit Phys. Lett.} {\refbf{#1}} }
\def\PR #1 {{\refit Phys. Rev. Lett.} {\refbf{#1}} }
\def\PRD #1 {{\refit Phys. Rev.} {\refbf D{#1}} }

% More nitpicking

\hyphenation{pre-print}
\hyphenation{quan-ti-za-tion}

%
% MATH (mostly)
%

% accent over:

\def\oonoo#1#2#3{\vbox{\ialign{##\crcr
	\hfil\hfil\hfil{$#3{#1}$}\hfil\crcr\noalign{\kern1pt\nointerlineskip}
	$#3{#2}$\crcr}}}
\def\oon#1#2{\mathchoice{\oonoo{#1}{#2}{\displaystyle}}
	{\oonoo{#1}{#2}{\textstyle}}{\oonoo{#1}{#2}{\scriptstyle}}
	{\oonoo{#1}{#2}{\scriptscriptstyle}}}
\def\dt#1{\oon{\hbox{\bf .}}{#1}}  
\def\ddt#1{\oon{\hbox{\bf .\kern-1pt.}}#1}    % À À   (see below)
\def\slap#1#2{\setbox0=\hbox{$#1{#2}$}
	#2\kern-\wd0{\hfuzz=1pt\hbox to\wd0{\hfil$#1{/}$\hfil}}}
\def\sla#1{\mathpalette\slap{#1}}                % slash: see Ö   below
\def\bop#1{\setbox0=\hbox{$#1M$}\mkern1.5mu
	\lower.02\ht0\vbox{\hrule height0pt depth.06\ht0
	\hbox{\vrule width.06\ht0 height.9\ht0 \kern.9\ht0
	\vrule width.06\ht0}\hrule height.06\ht0}\mkern1.5mu}
\def\bo{{\mathpalette\bop{}}}                        % box: see õ below
\def~{\widetilde} % tilde key; use Option-N for accent in text & math,
	% Option-' for sim, Option-0 for math space (¼)
\mathcode`\*="702A                  % * now always complex conjugate
\def\in{\relax\ifmmode\mathchar"3232\else{\refit in\/}\fi} % ã below 
\def\f#1#2{{\textstyle{#1\over#2}}}	   % fraction
\def\half{{\textstyle{1\over{\raise.1ex\hbox{$\scriptstyle{2}$}}}}}

% stick with math italic for cap Greek
\def\Gamma{\mathchar"0100}
\def\Delta{\mathchar"0101}
\def\Theta{\mathchar"0102}
\def\Lambda{\mathchar"0103}
\def\Xi{\mathchar"0104}
\def\Pi{\mathchar"0105}
\def\Sigma{\mathchar"0106}
\def\Upsilon{\mathchar"0107}
\def\Phi{\mathchar"0108}
\def\Psi{\mathchar"0109}
\def\Omega{\mathchar"010A}

\catcode128=13 \def €{\"A}                 % Option-u A
\catcode129=13 \def {\AA}                 % Option-A
\catcode130=13 \def '{\c}           	   % Option-C (cedilla)
\catcode131=13 \def ƒ{\'E}                   % Option-e E
\catcode132=13 \def "{\~N}                   % Option-n N
\catcode133=13 \def …{\"O}                 % Option-u O
\catcode134=13 \def †{\"U}                  % Option-u U
\catcode135=13 \def ‡{\'a}                  % Option-e a
\catcode136=13 \def ˆ{\`a}                   % Option-`  a
\catcode137=13 \def ‰{\^a}                 % Option-i a
\catcode138=13 \def Š{\"a}                 % Option-u a
\catcode139=13 \def ‹{\~a}                   % Option-n a
\catcode140=13 \def Œ{\alpha}            % Option-a
\catcode141=13 \def {\chi}                % Option-c
\catcode142=13 \def Ž{\'e}                   % Option-e e
\catcode143=13 \def {\`e}                    % Option-`  e
\catcode144=13 \def {\^e}                  % Option-i e
\catcode145=13 \def '{\"e}                % Option-u e
\catcode146=13 \def '{\'\i}                 % Option-e i
\catcode147=13 \def "{\`\i}                  % Option-`  i
\catcode148=13 \def "{\^\i}                % Option-i i
\catcode149=13 \def •{\"\i}                % Option-u i
\catcode150=13 \def –{\~n}                  % Option-n n
\catcode151=13 \def —{\'o}                 % Option-e o
\catcode152=13 \def ˜{\`o}                  % Option-`  o
\catcode153=13 \def ™{\^o}                % Option-i o
\catcode154=13 \def š{\"o}                 % Option-u o
\catcode155=13 \def ›{\~o}                  % Option-n o
\catcode156=13 \def œ{\'u}                  % Option-e u
\catcode157=13 \def {\`u}                  % Option-`  u
\catcode158=13 \def ž{\^u}                % Option-i u
\catcode159=13 \def Ÿ{\"u}                % Option-u u
\catcode160=13 \def  {\tau}               % Option-t
\catcode161=13 \mathchardef ¡="2203     % Option-* (TeX's usual eq. *)
\catcode162=13 \def ¢{\oplus}           % Option-4
\catcode163=13 \def £{\relax\ifmmode\to\else\itemize\fi} % Option-3
\catcode164=13 \def ¤{\subset}	  % Option-6
\catcode165=13 \def ¥{\infty}           % Option-8
\catcode166=13 \def ¦{\mp}                % Option-7
\catcode167=13 \def §{\sigma}           % Option-s
\catcode168=13 \def ¨{\rho}               % Option-r
\catcode169=13 \def ©{\gamma}         % Option-g
\catcode170=13 \def ª{\leftrightarrow} % Option-2 ; Option-E (acute) :
\catcode171=13 \def «{\relax\ifmmode\acute\else\expandafter\'\fi}
\catcode172=13 \def ¬{\relax\ifmmode\expandafter\ddt\else\expandafter\"\fi}
\catcode173=13 \def ­{\equiv}            % Option-= ; ^ Option-U (umlaudt)
\catcode174=13 \def ®{\approx}          % Option-"
\catcode175=13 \def ¯{\Omega}          % Option-O
\catcode176=13 \def °{\otimes}          % Option-5
\catcode177=13 \def ±{\ne}                 % Option-+
\catcode178=13 \def ²{\le}                   % Option-,
\catcode179=13 \def ³{\ge}                  % Option-.
\catcode180=13 \def ´{\upsilon}          % Option-y
\catcode181=13 \def µ{\mu}                % Option-m
\catcode182=13 \def ¶{\delta}             % Option-d
\catcode183=13 \def ·{\epsilon}          % Option-w
\catcode184=13 \def ¸{\Pi}                  % Option-P
\catcode185=13 \def ¹{\pi}                  % Option-p
\catcode186=13 \def º{\beta}               % Option-b
\catcode187=13 \def »{\partial}           % Option-9
\catcode188=13 \def ¼{\nobreak\ }       % Option-0
\catcode189=13 \def ½{\zeta}               % Option-z
\catcode190=13 \def ¾{\sim}                 % Option-'
\catcode191=13 \def ¿{\omega}           % Option-o
\catcode192=13 \def À{\dt}                     % Option-?
\catcode193=13 \def Á{\gets}                % Option-1
\catcode194=13 \def Â{\lambda}           % Option-l
\catcode195=13 \def Ã{\nu}                   % Option-v
\catcode196=13 \def Ä{\phi}                  % Option-f
\catcode197=13 \def Å{\xi}                     % Option-x
\catcode198=13 \def Æ{\psi}                  % Option-j
\catcode199=13 \def Ç{\int}                    % Option-\
\catcode200=13 \def È{\oint}                 % Option-|
\catcode201=13 \def É{\relax\ifmmode\cdot\else\vol\fi}    % Option-;
\catcode202=13 \def Ê{\relax\ifmmode\,\else\thinspace\fi}
\catcode203=13 \def Ë{\`A}                      % Option-`  A ; ^ Option-space
\catcode204=13 \def Ì{\~A}                      % Option-n A
\catcode205=13 \def Í{\~O}                      % Option-n O
\catcode206=13 \def Î{\Theta}              % Option-Q
\catcode207=13 \def Ï{\theta}               % Option-q; Option-- :
\catcode208=13 \def Ð{\relax\ifmmode\bar\else\expandafter\=\fi}
\catcode209=13 \def Ñ{\overline}             % Option-_
\catcode210=13 \def Ò{\langle}               % Option-[
\catcode211=13 \def Ó{\relax\ifmmode\{\else\ital\fi}      % Option-{
\catcode212=13 \def Ô{\rangle}               % Option-]
\catcode213=13 \def Õ{\}}                        % Option-}
\catcode214=13 \def Ö{\sla}                      % Option-/; Option-V :
\catcode215=13 \def ×{\relax\ifmmode\check\else\expandafter\v\fi}
\catcode216=13 \def Ø{\"y}                     % Option-u y
\catcode217=13 \def Ù{\"Y}  		    % Option-u Y
\catcode218=13 \def Ú{\Leftarrow}       % Option-!
\catcode219=13 \def Û{\Leftrightarrow}       % Option-@ ; Option-# :
\catcode220=13 \def Ü{\relax\ifmmode\Rightarrow\else\sect\fi}
\catcode221=13 \def Ý{\sum}                  % Option-$
\catcode222=13 \def Þ{\prod}                 % Option-%
\catcode223=13 \def ß{\widehat}              % Option-^
\catcode224=13 \def à{\pm}                     % Option-&
\catcode225=13 \def á{\nabla}                % Option-(
\catcode226=13 \def â{\quad}                 % Option-)
\catcode227=13 \def ã{\in}               	% Option-W
\catcode228=13 \def ä{\star}      	      % Option-R
\catcode229=13 \def å{\sqrt}                   % Option-M
\catcode230=13 \def æ{\^E}			% Option-i E
\catcode231=13 \def ç{\Upsilon}              % Option-Y
\catcode232=13 \def è{\"E}    	   	 % Option-u E
\catcode233=13 \def é{\`E}               	  % Option-`  E
\catcode234=13 \def ê{\Sigma}                % Option-S
\catcode235=13 \def ë{\Delta}                 % Option-D
\catcode236=13 \def ì{\Phi}                     % Option-F
\catcode237=13 \def í{\`I}        		   % Option-`  I
\catcode238=13 \def î{\iota}        	     % Option-H
\catcode239=13 \def ï{\Psi}                     % Option-J
\catcode240=13 \def ð{\times}                  % Option-K
\catcode241=13 \def ñ{\Lambda}             % Option-L
\catcode242=13 \def ò{\cdots}                % Option-:
\catcode243=13 \def ó{\^U}			% Option-i U
\catcode244=13 \def ô{\`U}    	              % Option-`  U
\catcode245=13 \def õ{\bo}                       % Option-B ; Option-I :
\catcode246=13 \def ö{\relax\ifmmode\hat\else\expandafter\^\fi}
\catcode247=13 \def÷{\relax\ifmmode\tilde\else\expandafter\~\fi}
\catcode248=13 \def ø{\ll}                         % Option-< ; ^ Option-N
\catcode249=13 \def ù{\gg}                       % Option-> 
\catcode250=13 \def ú{\eta}                      % Option-h 
\catcode251=13 \def û{\kappa}                  % Option-k 
\catcode252=13 \def ü{\half}     		 % Option-Z 
\catcode253=13 \def ý{\Gamma} 		% Option-G 
\catcode254=13 \def þ{\Xi}   			% Option-X ; Option-T : 
\catcode255=13 \def ÿ{\relax\ifmmode{}^{\dagger}{}\else\dag\fi}

% hat, check, tilde, and bar have been defined to work in text as well.

\def\ital#1Õ{{\it#1\/}}	     % for italics in text: see Ó above
\def\un#1{\relax\ifmmode\underline#1\else $\underline{\hbox{#1}}$
	\relax\fi}

	% for unitalicized
\def\roonoo#1#2#3{\vbox{\ialign{##\crcr
	\hfil{$#3{#1}$}\hfil\crcr\noalign{\kern1pt\nointerlineskip}
	$#3{#2}$\crcr}}}

	% accent under
\def\tdt#1{\oon{\hbox{\bf .\kern-1pt.\kern-1pt.}}#1}   % À À À
\def\({\eqno(}

%\def\refs{\sect{REFERENCES}\par\medskip \frenchspacing 
%	\parskip=0pt \refrm \baselineskip=1.23em plus 1pt
%	\def\ital##1Õ{{\refit##1\/}}}

% Young tableaux:  \upõ<a>{\õ<a>...\õ<b>}

\def\õ#1{
	\screwcount\num
	\num=1
	\screwdimen\downsy
	\downsy=-1.5ex
	\mkern-3.5mu
	õ
	\loop
	\ifnum\num<#1
	\llap{\raise\num\downsy\hbox{$õ$}}
	\advance\num by1
	\repeat}
\def\upõ#1#2{\screwcount\numup
	\numup=#1
	\advance\numup by-1
	\screwdimen\upsy
	\upsy=.75ex
	\mkern3.5mu
	\raise\numup\upsy\hbox{$#2$}}

%%%%%%%%%%%%%%%%%%%%%%%%%%%%%%%%%%%%%%%%

% postscript/pdf

\newcount\marknumber	\marknumber=1
\newcount\countdp \newcount\countwd \newcount\countht 

%
% for ordinary tex
%
\ifx\pdfoutput\undefined
\def\rgboo#1{}
\input epsf

\def\postscript#1{\special{" #1}}		% for dvips
\postscript{
	/bd {bind def} bind def
	/fsd {findfont exch scalefont def} bd
	/sms {setfont moveto show} bd
	/ms {moveto show} bd
	/pdfmark where		% printers ignore pdfmarks
	{pop} {userdict /pdfmark /cleartomark load put} ifelse
	[ /PageMode /UseOutlines		% bookmark window open
	/DOCVIEW pdfmark}
\def\bookmark#1#2{\postscript{		% #1=subheadings (if not 0)
	[ /Dest /MyDest\the\marknumber /View [ /XYZ null null null ] /DEST pdfmark
	[ /Title (#2) /Count #1 /Dest /MyDest\the\marknumber /OUT pdfmark}%
	\advance\marknumber by1}
\def\pdfklink#1#2{%
	\hskip-.25em\setbox0=\hbox{#1}%
		\countdp=\dp0 \countwd=\wd0 \countht=\ht0%
		\divide\countdp by65536 \divide\countwd by65536%
			\divide\countht by65536%
		\advance\countdp by1 \advance\countwd by1%
			\advance\countht by1%
		\def\linkdp{\the\countdp} \def\linkwd{\the\countwd}%
			\def\linkht{\the\countht}%
	\postscript{
		[ /Rect [ -1.5 -\linkdp.0 0\linkwd.0 0\linkht.5 ] 
		/Border [ 0 0 0 ]
		/Action << /Subtype /URI /URI (#2) >>
		/Subtype /Link
		/ANN pdfmark}{\rgb{1 0 0}{#1}}}
%
% for pdftex
%
\else
\def\rgboo#1{\pdfliteral{#1 rg #1 RG}}

\pdfcatalog{/PageMode /UseOutlines}		% bookmark window open
\def\bookmark#1#2{
	\pdfdest num \marknumber xyz
	\pdfoutline goto num \marknumber count #1 {#2}
	\advance\marknumber by1}
\def\pdfklink#1#2{%
	\noindent\pdfstartlink user
		{/Subtype /Link
		/Border [ 0 0 0 ]
		/A << /S /URI /URI (#2) >>}{\rgb{1 0 0}{#1}}%
	\pdfendlink}
\fi

\def\rgbo#1#2{\rgboo{#1}#2\rgboo{0 0 0}}
\def\rgb#1#2{\mark{#1}\rgbo{#1}{#2}\mark{0 0 0}}
\def\pdflink#1{\pdfklink{#1}{#1}}
\def\xxxlink#1{\pdfklink{[arXiv:#1]}{http://arXiv.org/abs/#1}}

\catcode`@=11

\def\wlog#1{}	% I don't care about new registers

% headers/footers

\def\makeheadline{\vbox to\z@{\vskip-36.5\p@
	\line{\vbox to8.5\p@{}\the\headline%
	\ifnum\pageno=\z@\rgboo{0 0 0}\else\rgboo{\topmark}\fi%
	}\vss}\nointerlineskip}
\headline={
	\ifnum\pageno=\z@
		\hfil
	\else
		\ifnum\pageno<\z@
			\ifodd\pageno
				\tenrm\romannumeral-\pageno\hfil\lefthead\hfil
			\else
				\tenrm\hfil\righthead\hfil\romannumeral-\pageno
			\fi
		\else
			\ifodd\pageno
				\tenrm\hfil\righthead\hfil\number\pageno
			\else
				\tenrm\number\pageno\hfil\lefthead\hfil
			\fi
		\fi
	\fi}

\catcode`@=12

\def\righthead{\hfil} \def\lefthead{\hfil}
\nopagenumbers

% divisions

%\def\bulletfill{\cleaders\hbox{$\mathsurround=0pt \mkern4mu
%	\raise.15em\hbox{$\bullet$} \mkern4mu$}\hfill}
\def\chrulefill{\rgb{1 0 0}{\hrulefill}}
\def\cdotfill{\rgb{1 0 0}{\dotfill}}
\newcount\area	\area=1
\newcount\cross	\cross=1
\def\volume#1\par{\newpage\noindent{\biggest{\rgb{1 .5 0}{#1}}}
	\par\nobreak\bigskip\medskip\area=0}
\def\chapskip{\par\ifnum\area=0\bigskip\medskip\goodbreak
	\else\newpage\fi}
\def\chapy#1{\area=1\cross=0
	\xdef\lefthead{\rgbo{1 0 .5}{#1}}\vbox{\biggerer\offinterlineskip
	\line{\chrulefill¼\hphantom{\lefthead}\chrulefill}
	\line{\chrulefill¼\lefthead\chrulefill}}\par\nobreak\medskip}
\def\chap#1\par{\chapskip\bookmark3{#1}\chapy{#1}}
\def\sectskip{\par\ifnum\cross=0\bigskip\medskip\goodbreak
	\else\newpage\fi}
\def\secty#1{\cross=1
	\xdef\righthead{\rgbo{1 0 1}{#1}}\vbox{\bigger\offinterlineskip
	\line{\cdotfill¼\hphantom{\righthead}\cdotfill}
	\line{\cdotfill¼\righthead\cdotfill}}\par\nobreak\medskip}
\def\sect#1 #2\par{\sectskip\bookmark{#1}{#2}\secty{#2}}
\def\subsectskip{\par\ifdim\lastskip<\medskipamount
	\bigskip\medskip\goodbreak\else\nobreak\fi}
\def\subsecty#1{\noindent{\sectfont{\rgbo{.5 0 1}{#1}}}\par\nobreak\medskip}
\def\subsect#1\par{\subsectskip\bookmark0{#1}\subsecty{#1}}
\long\def\x#1 #2\par{\hangindent2\parindent%
\mark{0 0 1}\rgboo{0 0 1}{\bf Exercise #1}\\#2%
\par\rgboo{0 0 0}\mark{0 0 0}}
\def\refs{\bigskip\noindent{\bf \rgbo{0 .5 1}{REFERENCES}}\par\nobreak\medskip
	\frenchspacing \parskip=0pt \refrm \baselineskip=1.23em plus 1pt
	\def\ital##1Õ{{\refit##1\/}}}
\long\def\twocolumn#1#2{\hbox to\hsize{\vtop{\hsize=2.9in#1}
	\hfil\vtop{\hsize=2.9in #2}}}

% fonts

\twelvepoint
\font\bigger=cmbx12 \sca2
\font\biggerer=cmb10 \sca5
%\font\biggest=cmssdc10 scaled 3583
%\font\biggest=cmssdc10 scaled 4250
\font\biggest=cmssdc10 scaled 3700
%\font\subtitlefont=cmbxti10 scaled 3583
 \sca5

 \sca3

% symbols

\def Ü{\relax\ifmmode\Rightarrow\else\expandafter\subsect\fi}
\def Û{\relax\ifmmode\Leftrightarrow\else\expandafter\sect\fi}
\def Ú{\relax\ifmmode\Leftarrow\else\expandafter\chap\fi}

\def\itemize#1 {\item{\bf#1}}
\def\itemizze#1 {\itemitem{\bf#1}}
\def\itemutem{\par\indent\indent \hangindent3\parindent \textindent}
\def\itemizzze#1 {\itemutem{\bf#1}}
\def ª{\relax\ifmmode\leftrightarrow\else\itemizze\fi}
\def Á{\relax\ifmmode\gets\else\itemizzze\fi}

\def\¢{\ominus}

\def\Ä{\varphi}  \def\¿{\varpi}	\def\Ï{\vartheta}

\def ò{\relax\ifmmode\cdots\else\dotfill\fi}

\chardef\slo="1C

% boxes drawn around phrases or paragraphs

\def\cvrule{\rgbo{0 .5 1}{\vrule}}
\def\chrule{\rgbo{0 .5 1}{\hrule}}
\def\boxit#1{\leavevmode\thinspace\hbox{\cvrule\vtop{\vbox{\chrule%
	\vskip3pt\kern1pt\hbox{\vphantom{\bf/}\thinspace\thinspace%
	{\bf#1}\thinspace\thinspace}}\kern1pt\vskip3pt\chrule}\cvrule}%
	\thinspace}
\def\Boxit#1{\noindent\vbox{\chrule\hbox{\cvrule\kern3pt\vbox{
	\advance\hsize-7pt\vskip-\parskip\kern3pt\bf#1
	\hbox{\vrule height0pt depth\dp\strutbox width0pt}
	\kern3pt}\kern3pt\cvrule}\chrule}}

% boxes around equations

          % inside $$'s
   % outside $$'s

% other

\def\today{\ifcase\month\or
 January\or February\or March\or April\or May\or June\or July\or
 August\or September\or October\or November\or December\fi
 \space\number\day, \number\year}

\parindent=20pt
\newskip\normalparskip	\normalparskip=.7\medskipamount
\parskip=\normalparskip	% space between paragraphs

%%%%%%%%%%%%%%%%%%%%%%%%%%%%%%%%%%%%%%%%

% Some stupid little things that have to go at the end:
% make |,<,> OK in text

\catcode`\|=\active \catcode`\<=\active \catcode`\>=\active 
\def|{\relax\ifmmode\delimiter"026A30C \else$\mathchar"026A$\fi}
\def<{\relax\ifmmode\mathchar"313C \else$\mathchar"313C$\fi}
\def>{\relax\ifmmode\mathchar"313E \else$\mathchar"313E$\fi}

%%%%%%%%%%%%%%%%%%%%%%%%%%%%%%%%%%%%%%%%

% PAPER:
%	\paper
%
%	"title"
%
%	"authors"
%
%	"preprint number"
%
%	"date"
%
%	"abstract"
%
%	"text"
%
%	\bye

\def\thetitle#1#2#3#4#5{
 \def\titlefont{\biggest} \font\footrm=cmr10 \font\footit=cmti10
  \twelverm
	{\hbox to\hsize{#4 \hfill YITP-SB-#3}}\par
	\vskip.8in minus.1in {\center\baselineskip=2.2\normalbaselineskip
 {\titlefont #1}\par}{\center\baselineskip=\normalbaselineskip
 \vskip.5in minus.2in #2
	\vskip1.4in minus1.2in {\twelvebf ABSTRACT}\par}
 \vskip.1in\par
 \narrower\par#5\par\unnarrower\vskip3.5in minus3.3in\eject}
\def\paper\par#1\par#2\par#3\par#4\par#5\par{
	\thetitle{#1}{#2}{#3}{#4}{#5}} 
\def\author#1#2{#1 \vskip.1in {\twelveit #2}\vskip.1in}
\def\YITP{C. N. Yang Institute for Theoretical Physics\\
	State University of New York, Stony Brook, NY 11794-3840}
\def\WS{W. Siegel\footnote{$*$}{% e.g.,\author\WS\YITP
	\pdflink{mailto:siegel@insti.physics.sunysb.edu}\\
	\pdfklink{http://insti.physics.sunysb.edu/\~{}siegel/plan.html}
	{http://insti.physics.sunysb.edu/\noexpand~siegel/plan.html}}}

%%%%%%%%%%%%%%%%%%%%%%%%%%%%%%%%%%%%%%%%

\pageno=0

\paper

{\rgb{1 0.5 0}{Yang-Mills by dimensionally reducing Chern-Simons}}

\author\WS\YITP

11-20

June 27, 2011

We derive the usual first-order form of the Yang-Mills action in arbitrary dimensions by dimensional reduction from a Chern-Simons-like action.  The antisymmetric tensor auxiliary field of the first-order action appears as a gauge field for the extra dimensions.  The higher-dimensional geometry was introduced in our previous paper by adding dimensions ``dual" to spin, as suggested by the superstring's affine Lie algebra.

\pageno=2

ÜChern-Simons analogs

We consider a class of theories related to Chern-Simons, but defined in more than three dimensions, and dependent on the geometry.  The theories are defined as usual in terms of Yang-Mills covariant derivatives $á_A=d_A+iA_A$, but the free derivatives $d_A$ are nonabelian.  Dividing up the derivatives as $A=(Œ,i)$,
$$ [d_A,d_BÕ = T_{AB}{}^C d_C :ââ[d_Œ,d_ºÕ = T_{Œº}{}^i d_i,ââ[d_Œ,d_iÕ = [d_i,d_jÕ = 0 $$
The field strengths are
$$ F_{AB} = d_{[A}A_{B)} -T_{AB}{}^C A_C +iA_{[A}A_{B)} $$
Defining as usual the Chern-Simons form (but taking into account the torsion)
$$ X_{ABC} = üA_{[A}d_B A_{C)} -\f14 A_{[A}T_{BC)}{}^D A_D +\f13 iA_{[A}A_B A_{C)} $$
the Lagrangian takes the form (appearing as its trace in the action)
$$ L = üa^{Œºi}X_{Œºi} $$
for some constant tensor $a$ invariant under the desired symmetries.

All such actions have an important difference from standard Chern-Simons terms:  The torsion introduces a contribution
$$ -üú^{ij}A_i A_j,ââú^{ij} ­ \f14 a^{Œº(i}T_{Œº}{}^{j)} $$
We assume $ú$ is invertible, which allows $A_i$ to be removed from the action as an auxiliary field by its field equation.  The action thus resembles the first-order form of a standard Yang-Mills action:
$$ L =  (-üú^{ij}A_i A_j +A_i öF^i) -üa^{Œºi}A_Œ d_i A_º,ââ
	öF^i ­ üa^{Œºi}(d_{[Œ}A_{º)} +iA_{[Œ}A_{º)}) $$
But $öF$ is not covariant; in fact the auxiliary field equation is $ú^{ij}A_j=öF^i$.

There are several nontrivial examples of such actions in the literature.  One is the minimal supersymmetrization of the usual 3D Chern-Simons form [1]:  In that case $d_Œ$ are the usual 2 supersymmetry derivatives and $d_i$ the 3 translations, while $T$ and $a$ are both $©$ matrices.  The auxiliary field equation is just the usual conventional constraint determining the vector gauge field in terms of the spinor.  A similar expression yields the action for 4D N=1 super Yang-Mills (with $Œ$ a 4-spinor index), if the representation-preserving constraints are imposed by hand [2].  Another example is the action for maximally supersymmetric Yang-Mills in 4D N=3 harmonic superspace¼[3]:  In that case the 2 $d_Œ$ and 1 $d_i$ are the 3 lowering operators of SU(3)/U(1)$^2$.  Unlike the previous cases, not all the derivatives of the space appear in that Lagrangian.

ÜDimensional reduction

In previous papers [4] we derived the affine Lie algebra of the superstring, containing the generalization of the superparticle's spinor and vector derivatives, but also a spinor ``dual" to that spinor derivative.  The spinor field strength of super Yang-Mills appeared as a ``gauge field" to the new spinor derivative, in the same way as the usual spinor and vector gauge superfields did for the other two derivatives.

In a recent paper [5] we considered the natural extension of this affine Lie algebra to include a derivative for which the antisymmetric tensor field strength is the gauge field.  It appeared as a necessary consequence of including spin operators in the algebra, as their dual.  (``Prepotentials" appeared as gauge fields for the spin.)

We then applied this algebra directly to super Yang-Mills and supergravity, without direct reference to strings.  Superspace was derived by a combination of isotropy constraints, imposed in terms of gauge covariant derivatives (for the spin derivatives), and dimensional reduction, imposed in terms of dual symmetry generators.  The former eliminated also the corresponding gauge fields, in covariant gauges, while the latter kept the corresponding gauge fields, but as field strengths (as in conventional forms of dimensional reduction).

In this paper we consider only bosonic Yang-Mills, so we limit our algebra of derivatives to only the usual translations and the derivatives dual to spin.  The algebra is a special case of that considered in the previous section:
$$ [d_a,d_b] = d_{ab},ââ[d_{ab},d_c] = [d_{ab},d_{cd}] = 0 $$
with $Œ=a$, $i=ab$.  We now have
$$ T_{a,b}{}^{cd} = ü¶_{[a}^c ¶_{b]}^d,ââa^{a,b,cd} = ú^{a[c}ú^{d]b},ââ
	ú^{ab,cd} = üú^{a[c}ú^{d]b} $$
$$ öF^{ab} = d^{[a}A^{b]} +iA^{[a}A^{b]} $$

In our previous paper we did not consider an action for the theory before dimensional reduction, although this is generally how this reduction is applied, especially in supersymmetric theories.  We now see that the Chern-Simons-like action above is the one suited for this purpose:  (1) It directly identifies the gauge field $A_{ab}$ for the dual spin coordinates as the usual field strength $öF$, which is covariant upon reduction because the noncovariant $d_{ab}$ term in its transformation law dies.  (2) It gives the usual first-order Yang-Mills action upon reduction, since the extra $A_a d^{ab} A_b$ term drops out.

In D=4, the extra dimensions (and thus the extra field) can be restricted to be self-dual.

ÜConclusions

We have shown a natural way of introducing first-order formalisms, where the auxiliary field is originally nontrivial, appearing as a gauge field.  This field (and the corresponding extra dimensions) is also suggested by an algebraic analysis of the superstring.  This approach might have advantages similar to those of the manifestly T-dual formulation of the actions for low energy states of strings, which is treated as dimensional reduction from twice the coordinates [6].

Generalization to higher spins (e.g., gravity) and supersymmetry should be considered.  For example, the usual auxiliary fields of 4D N=1 and 2 super Yang-Mills (dimension-2 scalars in the adjoint of U(1) and SU(2) R-symmetry, respectively) would appear as the gauge fields for extra dimensions dual to R-symmetry (instead of spin), since they appear in generalized d'Alembertians as the R-symmetry analog of the relativistic Pauli term.  Another possible application would be to find a first-order formalism for string field theory, with on-shell field strengths built into the formalism.  A first-quantized formulation of the approach would be a first step.

ÜAcknowledgment

This work is supported in part by National Science Foundation Grant No.¼PHY-0969739.

\refs

£1 %\cite{Nishino:1991sr}
%\bibitem{Nishino:1991sr}
  H. Nishino and S.J. Gates, Jr.,
  %``Chern-Simons theories with supersymmetries in three-dimensions,''
  ÓInt. J. Mod. Phys.Õ ÉA8 (1993) 3371.

£2 %\cite{Siegel:1995px}
%\bibitem{Siegel:1995px}
  W. Siegel,
  %``Curved extended superspace from Yang-Mills theory a la strings,''
  \PRD 53 (1996)  3324
  \xxxlink{hep-th/9510150}.

£3 %\cite{Galperin:1984bu}
%\bibitem{Galperin:1984bu}
  A. Galperin, E. Ivanov, S. Kalitsyn, V. Ogievetsky, and E. Sokatchev,
  %``Unconstrained Off-Shell N=3 Supersymmetric Yang-Mills Theory,''
  ÓClass. Quant. Grav.Õ É2 (1985) 155,
  %\cite{Galperin:1985uw}
%\bibitem{Galperin:1985uw}
%  A.~Galperin, E.~Ivanov, S.~Kalitsyn, V.~Ogievetsky, E.~Sokatchev,
  %``N = 3 Supersymmetric Gauge Theory,''
  \PL 151B (1985) 215.

£4 W. Siegel, Covariant approach to superstrings, ã Symposium on anomalies, geometry and topology, eds. W.A. Bardeen and A.R. White (World Scientific, Singapore, 1985) p. 348;\\
 %\cite{Siegel:1985xj}
%\bibitem{Siegel:1985xj}
% W. Siegel,
  %``Classical Superstring Mechanics,''
  \NP 263 (1986) 93.

£5 %\cite{Siegel:2011sy}
%\bibitem{Siegel:2011sy}
  W. Siegel,
  New superspaces/algebras for superparticles/strings,
  \xxxlink{1106.1585} [hep-th].

£6 %\cite{Siegel:1993th}
%\bibitem{Siegel:1993th}
  W. Siegel,
  %``Superspace duality in low-energy superstrings,''
  \PRD 48 (1993) 2826
  \xxxlink{hep-th/9305073};
  %\cite{Siegel:1993bj}
%\bibitem{Siegel:1993bj}
%  W.~Siegel,
  Manifest duality in low-energy superstrings,
  ã Strings '93, eds. M.B. Halpern, G. Rivlis, and A. Sevrin (World Scientific, Singapore, 1993) p. 353
  \xxxlink{hep-th/9308133}.

\bye